\documentclass[aps,amsmath,amssymb,prd,superscriptaddress,twocolumn]{revtex4-1}

\usepackage{graphicx}
\usepackage{dcolumn}
\usepackage{bm}
\usepackage{color}
\usepackage{amsmath}
\usepackage[colorlinks=true,pdfstartview=FitV,linkcolor=blue,citecolor=blue,urlcolor=blue]{hyperref}

\everymath{\displaystyle}
\begin{document}

\newcommand{\up}[1]{$^{#1}$}
\newcommand{\down}[1]{$_{#1}$}
\newcommand{\powero}[1]{\mbox{10$^{#1}$}}
\newcommand{\powert}[2]{\mbox{#2$\times$10$^{#1}$}}

\newcommand{\evm}{\mbox{eV\,$c^{-2}$}}
\newcommand{\gevm}{\mbox{GeV\,$c^{-2}$}}
\newcommand{\pgd}{\mbox{g$^{-1}$\,d$^{-1}$}}
\newcommand{\mv}{\mbox{$m_V$}}
\newcommand{\um}{\mbox{$\mu$m}}
\newcommand{\spix}{\mbox{$\sigma_{\rm pix}$}}
\newcommand{\ame}{\mbox{$^{241}$Am}}
\newcommand{\cob}{\mbox{$^{57}$Co}}

\title{Measurement of low energy ionization signals from Compton scattering\\ in a CCD dark matter detector}

\author{K.~Ramanathan}
\affiliation{Kavli Institute for Cosmological Physics and The Enrico Fermi Institute, The University of Chicago, Chicago, IL, United States}

\author{A.~Kavner}
\affiliation{University of Michigan, Department of Physics, Ann Arbor, MI, United States}    

\author{A.E.~Chavarria}
\affiliation{Kavli Institute for Cosmological Physics and The Enrico Fermi Institute, The University of Chicago, Chicago, IL, United States}

\author{P.~Privitera}
\affiliation{Kavli Institute for Cosmological Physics and The Enrico Fermi Institute, The University of Chicago, Chicago, IL, United States}

\author{D.~Amidei}
\affiliation{University of Michigan, Department of Physics, Ann Arbor, MI, United States}

\author{T.-L.~Chou}
\affiliation{Kavli Institute for Cosmological Physics and The Enrico Fermi Institute, The University of Chicago, Chicago, IL, United States}

\author{A.~Matalon}
\affiliation{Kavli Institute for Cosmological Physics and The Enrico Fermi Institute, The University of Chicago, Chicago, IL, United States}

\author{R.~Thomas}
\affiliation{Kavli Institute for Cosmological Physics and The Enrico Fermi Institute, The University of Chicago, Chicago, IL, United States}

\author{J.~Estrada}
\affiliation{Fermi National Accelerator Laboratory, Batavia, IL, United States}

\author{J.~Tiffenberg}
\affiliation{Fermi National Accelerator Laboratory, Batavia, IL, United States}

\author{J.~Molina}
\affiliation{Facultad de Ingenier\'{\i}a, Universidad Nacional de Asunci\'on, Asuncion, Paraguay}

\date{\today}

\begin{abstract}
An important source of background in direct searches for low-mass dark matter particles are the energy deposits by small-angle scattering of environmental $\gamma$ rays.
We report detailed measurements of low-energy spectra from Compton scattering of $\gamma$ rays in the bulk silicon of a charge-coupled device (CCD).
Electron recoils produced by $\gamma$ rays from \cob\ and \ame\ radioactive sources are measured between 60\,eV and 4\,keV.
The observed spectra agree qualitatively with theoretical predictions, and characteristic spectral features associated with the atomic structure of the silicon target are accurately measured for the first time.
A theoretically-motivated parametrization of the data that describes the Compton spectrum at low energies for any incident $\gamma$-ray flux is derived.
The result is directly applicable to background estimations for low-mass dark matter direct-detection experiments based on silicon detectors, in particular for the DAMIC experiment down to its current energy threshold.
\end{abstract}


\maketitle

\section{Introduction}
\label{sec:intro}

Solid-state ionization detectors have been proposed for next-generation direct searches for dark matter~\cite{Aguilar-Arevalo:2016ndq, Agnese:2016cpb}. Thanks to their very low noise and the small band gap of the semiconductor target, these detectors are most sensitive to low-mass ($<$10\,GeV/$\rm{c}^2$) dark matter particles by their interactions with nuclei~\cite{Kolb:1990vq, *Griest:2000kj, *Zurek:2013wia} or electrons~\cite{Essig:2015cda, *Hochberg:2016sqx, *Bloch:2016sjj} in the target. At the low energies of interest for these searches, which correspond to ionization signals in the range 2--1000\,$e^-$, the dominant background from environmental radiation are the low-energy electron recoils from small-angle Compton scattering of external $\gamma$ rays, whose flux is generally orders of magnitude higher than fast neutrons, the other possible external source of background in the bulk of the target in this energy range.

In the presence of an irreducible electron-recoil background from Compton scattering, a potential signal from interactions of dark matter particles can only be identified by its spectrum. Therefore, a complete understanding of the low-energy spectral features of the ionization signals from Compton scattering is required for the success of low-mass dark matter searches.

In this paper, we report the measured spectra from Compton scattering of $\gamma$ rays above an energy of 60\,eV, corresponding to ionization signals of 15\,$e^-$. The results were obtained by exposing a high-resistivity fully depleted CCD~\cite{1185186} developed in the R\&D efforts of DAMIC~\cite{Aguilar-Arevalo:2016ndq, Aguilar-Arevalo:2016zop} to $\gamma$ rays from radioactive sources.
The measurements are found to be in fair agreement with the theoretical expectation, and we derive an improved phenomenological parametrization that more accurately describes Compton spectra in the regime of atomic binding, which can be used to predict the background from Compton scattering of $\gamma$ rays at low energies. 

\section{Compton scattering}
\label{sec:theory}

Compton scattering~\cite{PhysRev.21.483} is an electromagnetic process where an incident photon transfers some of its energy to an electron, and is then deflected from its original direction. For an interaction with a free electron at rest, the energy of a scattered photon ($E_s$) depends on the energy of the incident photon ($E_\gamma$), the mass of the electron ($m$) and the scattering angle ($\theta$) as

\begin{equation*}
	E_{s} = \frac{E_\gamma}{1 + \frac{E_\gamma}{mc^2}(1-\cos\theta)},
	\label{eq:compton}
\end{equation*}
with the differential cross section given by the well-known Klein-Nishina formula~\cite{Klein1929}. The maximum energy transferred to the electron $E=E_\gamma-E_s$ occurs when the $\gamma$ ray backscatters, i.e., when $\theta$$=$$\pi$, giving rise to a spectral feature known as the Compton edge.

The spectrum of energy deposited by a single $\gamma$-ray interaction in a target is a continuum from zero up to the Compton edge. However, the electrons in the target are bound in atomic shells with non-zero momentum, which lead to deviations from the Klein-Nishina formula and give rise to observable distortions in the spectrum, including the well-known softening or ``Doppler broadening" of the Compton edge~\cite{PhysRev.33.643}.

A straightforward modification of the Klein-Nishina formula for bound electrons is the Relativistic Impulse Approximation (IA)~\cite{PhysRevA.26.3325}, where each electron in an atomic shell is treated as a free electron with a constrained momentum distribution derived from the bound-state wave function. For low energy and momentum transfers, the differential cross section for a photon scattering with an atomic electron with quantum numbers $n$ and $l$ reduces to

\begin{equation}
\begin{aligned}
	\left. \frac{d\sigma}{dE} \right |_{nl} &= {\cal C} \int_{-1}^{1} \frac{(1-\delta)(1+\cos^2\theta)+\delta^2}{|\vec{q}|} J_{nl}(p_z)\,d\cos\theta\\
	p_z &= \frac{(E_\gamma/c)(1-\delta)(1-\cos\theta) - \delta mc}{|\vec{q}|}\\
	 |\vec{q}| &= \sqrt{2(1-\delta)(1-\cos\theta)+\delta^2} .
	\label{eq:doublediff}
\end{aligned}
\end{equation}
The expression above is only valid for $E$$\ge$$E_{nl}$, the target electron's binding energy.
Otherwise, $d\sigma/dE |_{nl}$$=$0, as the minimum energy that the photon can lose in an interaction is that required to free the target electron from the atom.
Here, we have introduced $\delta$$=$$E/E_\gamma$, which is $\ll$1 for the energies of interest, and grouped the constant terms in front of the expression as ${\cal C}$$=$$\pi r_0^2 mc/E_\gamma$, where $r_0$ is the classical electron radius.
The functions $J_{nl}(p_z)$ are the Compton profiles, which encode the momentum distribution of the target electron before the collision, and $p_z$ is the projection of the momentum of the electron on the scattering vector $\vec{q}$.
The integral in Eq.~\ref{eq:doublediff} can only be evaluated numerically. Tabulated data for $J_{nl}(p_z)$ in units of 1/($\alpha mc$) for different elements (listed by atomic number $Z$) can be found in Ref.~\cite{Biggs1975201}.

\begin{figure}[t!]
	\centering
	\includegraphics[width=0.482\textwidth]{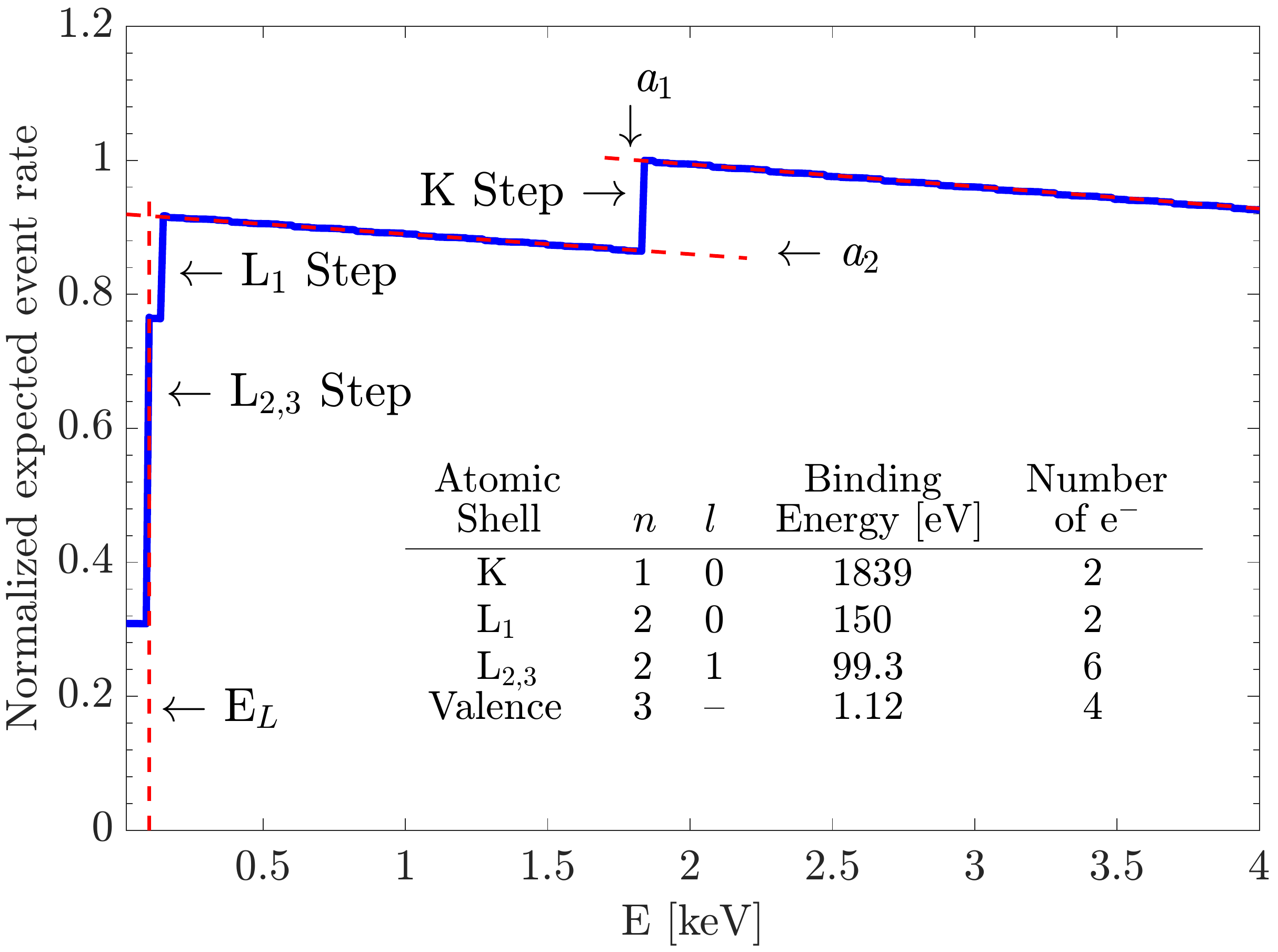}
	\caption{Computed low-energy spectrum from Compton scattering of 122 keV $\gamma$ rays in silicon. The observed steps occur at the binding energies of the different atomic shells, as given in the inset~\cite{nistxray}. The spectrum was normalized so that its value is one on the right-hand side of the K step. Parameters $a_1$, $a_2$ and $E_L$ of the general model presented in Section~\ref{sec:model} are labeled.}
	\label{fig:Biggs}
\end{figure}

Figure~\ref{fig:Biggs} shows the computed spectrum for a silicon target exposed to 122\,keV $\gamma$ rays, where we added $d\sigma/dE |_{nl}$ over all atomic electrons.
A series of steps are observed at low energies, corresponding to the atomic shells of the target, which arise from the condition that  $d\sigma/dE |_{nl}$$=$0 for $E$$<$$E_{nl}$.
At threshold, the freed electron has negligible kinetic energy and the energy deposited comes from the refilling of the atomic vacancy by the emission of secondary Auger electrons and fluorescence x rays.
An approximate estimate of the slope of the spectrum between the steps can be obtained from the Klein-Nishina formula, whose solution for $\delta$$\ll$1 reduces to $d\sigma/dE \propto 1 - (mc^2/E_\gamma^2)E$.

The experimental observation of these spectral features, which are of particular relevance to the understanding of the radioactive backgrounds for low-mass dark matter searches, has never before been reported in the literature.

\section{Experimental setup}
\label{sec:setup}

\begin{figure}[t!]
	\centering
	\includegraphics[width=0.482\textwidth]{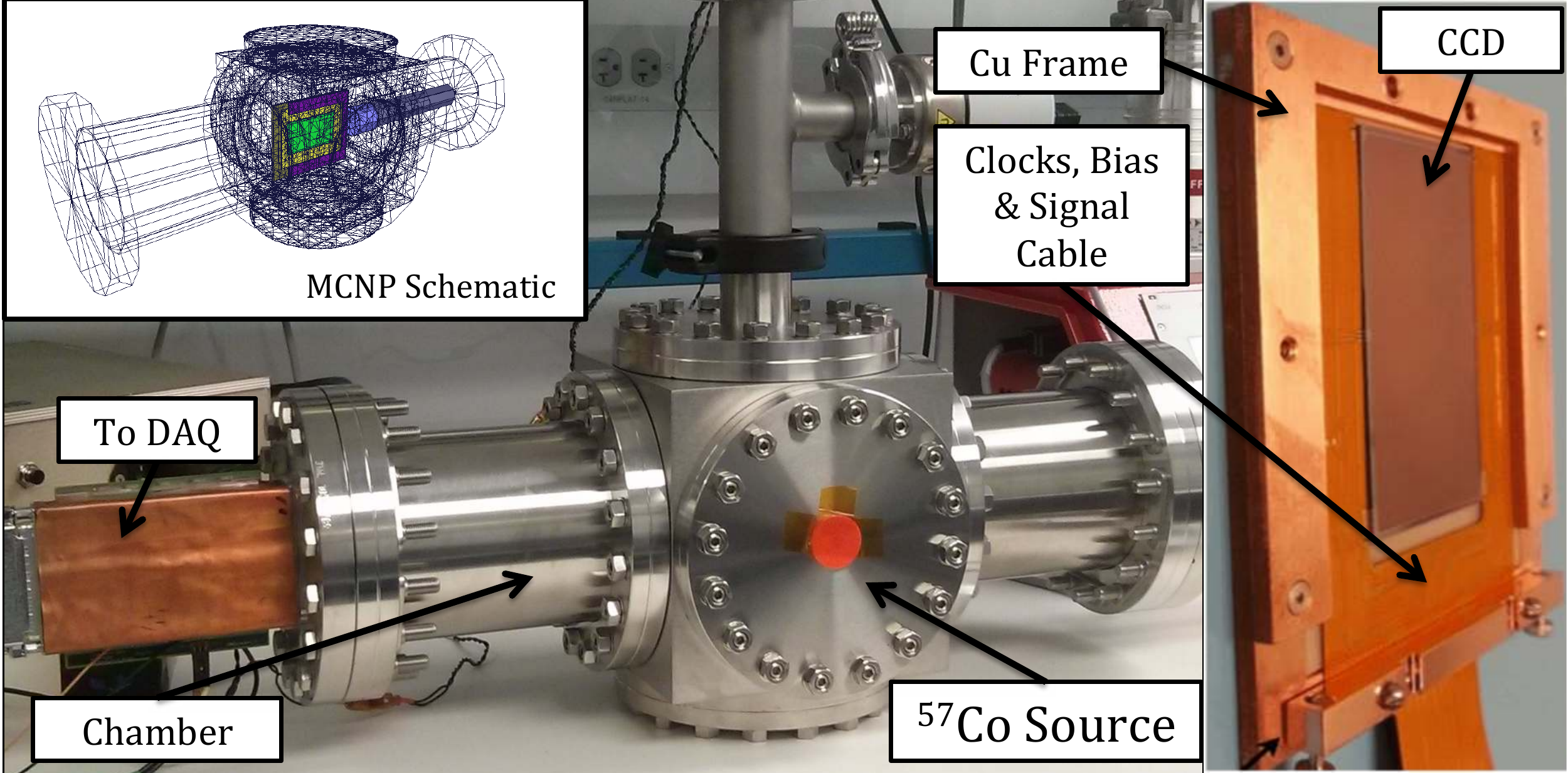}
	\caption{ {\it Left}: Experimental setup at The University of Chicago, showing the \cob\ radioactive source installed on the front flange of the stainless steel vacuum chamber. {\it Inset}: MCNP model of the experimental setup with the CCD location inside the vacuum chamber shown in green. {\it Right}: Packaged 8\,Mpixel CCD in its copper module.}
	\label{fig:SetupImage}
\end{figure}

The setup used for this measurement is shown in Fig.~\ref{fig:SetupImage}. We employed an 8\,Mpixel CCD (pixel size 15$\times$15\,$\mu$m$^2$) with an active area of 18.8\,cm$^2$, a thickness of 500\,$\mu$m and a mass of 2.2\,g. The response of this device to ionizing radiation has been previously characterized~\cite{Chavarria:2016xsi}. The CCD was installed in a stainless-steel vacuum chamber ($10^{-6}$\,mbar) and cooled to a nominal operating temperature of 130\,K. The voltage biases, clocks and video signals required for the CCD operation were serviced by a Kapton flex cable wire bonded to the CCD. The silicon substrate was fully depleted by an external bias, with no regions of partial or incomplete charge collection that may hinder the energy response of the device~\cite{Aguilar-Arevalo:2016ndq, 1185186}. The CCD was controlled and read out by commercial CCD electronics (Astronomical Research Cameras, Inc.). The pixel noise achieved with this system was 1.86$\pm$0.02\,$e^-$, equivalent to 7.0$\pm$0.1\,eV  (on average, 3.8\,eV are required to produce a free charge carrier in silicon at 130\,K~\cite{4326950}).

Compton-scattered electrons produce charge in the bulk of the device by ionization, with the number of charge carriers being proportional to the kinetic energy of the electron.
The charge carriers are drifted along the direction of the electric field ($z$ axis) and collected on the pixel array.
Because of thermal motion, the ionized charge diffuses transversely with respect to the electric field direction as it is drifted, with a spatial variance ($\sigma_{xy}^2$) that is proportional to the carrier transit time.
Hence, there is a positive correlation between the lateral diffusion ($\sigma_{xy}$) of the collected charge on the pixel array and the depth of the interaction ($z$)~\cite{Aguilar-Arevalo:2016ndq,1352164,*CeaseDiff}.

The CCD images contain a two-dimensional (2D) stacked history (projected on the $x$-$y$ plane) of all ionization produced throughout an exposure, where each image pixel value is proportional to the collected number of charge carriers.
Data for this analysis were acquired in two modes: i) standard 1$\times$1 binning, where each pixel was read out individually for maximum spatial resolution, and ii) 4$\times$4 binning, where by the appropriate clocking of the device the charge collected in groups of 4$\times$4 pixels was read out in a single measurement.
Since readout noise is introduced only once on the larger charge signal given by the sum of the group of pixels, a better signal-to-noise ratio is obtained with 4$\times$4 binning at the expense of a worsened spatial resolution.
For details on the readout modes of DAMIC CCDs see Ref.~\cite{Aguilar-Arevalo:2016ndq}.

Radioactive sources of \ame\ and \cob\ with properties listed in Table~\ref{tab:sources} were installed on the front flange of the vessel (Fig.~\ref{fig:SetupImage}). These isotopes were chosen as they provide a few intense $\gamma$-ray lines (60\,keV, 122\,keV and 136\,keV) of relatively low energy for which Compton scattering is the dominant interaction in the silicon target. Lower energy $\gamma$ rays are preferred for statistical considerations, as the Compton spectrum is compressed toward lower energies with a larger fraction of interactions close to atomic binding energies. However, if the $\gamma$-ray energy is too low (e.g., 14\,keV and 26\,keV in Table~\ref{tab:sources}) photoelectric absorption dominates with no observable Compton spectrum. Furthermore, lower-energy $\gamma$ rays lead to shorter-range electron recoils, mitigating surface effects by minimizing both the number of electrons that escape the CCD without depositing their full energy and the flux of degraded-energy electrons arising from the surfaces of materials surrounding the device.

\begin{table}[b!]
	\caption{\label{tab:sources}Summary of the radioactive sources used in this experiment. The energies ($E_\gamma$) and intensities of the relevant $\gamma$-ray lines are presented. Values from Ref.~\cite{be2004table, *be2010table}.}
	\begin{ruledtabular}
		\begin{tabular}{ccccc}
			Source & Activity & Half-Life & $E_\gamma$ & Intensity\\
			& [$\mu$Ci]     & [y] 	     & [keV]	    & [per 100 decays]\\
			\hline
			\cob & 8.7 & 0.745\,(1) & 14.4130 (3)  & 9.2\,(2)\\
			&	   &           & 122.0607\,(1) & 85.51\,(6)\\
			&	   &           & 136.4736\,(3) & 10.7\,(2)\\
			\ame & 22  & 432.6\,(6) & 26.3446\,(2)    & 2.31\,(8)\\
			&	   &           & 59.5409\,(1)    & 35.9\,(2) \\				
		\end{tabular}
	\end{ruledtabular}
\end{table}

\section{Data sets}
\label{sec:dataset}

The eight sets of images used for this analysis are summarized in Table~\ref{tab:dataset}.
For each radioactive source, we acquired data with 1$\times$1 and 4$\times$4 binning (Section~\ref{sec:setup}), each followed by a background run with the source removed.
The 1$\times$1 data were used to confirm the presence of a dominant bulk signal from Compton scattering relative to surface backgrounds.
Thus, the data were acquired at low substrate bias ($V_{\rm sub}$) to increase lateral charge diffusion, and offer maximum spatial resolution for the precise reconstruction of the three-dimensional (3D) location of electron recoils.
The 4$\times$4 data were used to perform spectroscopy at the lowest energies.
Hence, the data were acquired with a high $V_{\rm sub}$ so that most of the charge from an interaction was collected in a single 4$\times$4 pixel group, minimizing the contribution from readout noise in the charge measurement.
Finally, the background data were acquired to characterize and monitor the contribution from electronic noise and environmental backgrounds to the source runs.

\begin{table}[b!]
	\caption{\label{tab:dataset} Summary of the data sets used in the analysis. The event density was estimated in the 1--5\,keV range after the masking procedure outlined in Section~\ref{sec:event}.}
	\begin{ruledtabular}
		\begin{tabular}{cccccc}
			Binning & Source &  $V_{\rm sub}$ & $N$ images & Exposure &  Event density \\
				    &  		   & 	[V]	     &	     &   [s]	&  [keV$^{-1}$]  \\
			\hline
		 1$\times$1 & \cob\  &45 & 1898  &986& 3.5$\times$$10^4$  \\
				 & None & 45 & 1326 &986& 4.3$\times$$10^3$  \\
	      			  & \ame\  & 45 & 971 &490& 4.7$\times$$10^4$  \\
	       			& None & 45 &1235  &490&  2.4$\times$$10^3$\\
		4$\times$4& \cob\  & 127 &1815  &39.8& 2.3$\times$$10^5$  \\
				 & None & 127 & 2060  &39.8& 2.6$\times$$10^2$ \\
				 & \ame\  & 127 & 9828  &39.8& 2.5$\times$$10^5$ \\
				  & None & 127 & 10267 &39.8& 1.1$\times$$10^3$ \\
		\end{tabular}
	\end{ruledtabular}
\end{table}

A simulated data set for every detector-source configuration was produced with MCNPX (Monte Carlo N-Particle eXtended; v.2.7.0) particle transport code~\cite{mcnp}.
Full photon and electron transport was enabled using the corrected MCPLIB84 library to properly account for Doppler broadening~\cite{mcnpCorr}.
The geometry and material specification of the setup was accurately reproduced (inset of Fig.~\ref{fig:SetupImage}).
The production of charge carriers and their diffusion as they drift to the CCD pixel array were simulated with a dedicated Monte Carlo code. 
For each energy deposit, we simulated a 2D Gaussian distribution of charge on the pixel array, with a total number of charge carriers that is proportional to the deposited energy and a standard deviation ($\sigma_{xy}$) that is related to the depth ($z$) of the energy deposit by the charge diffusion model presented and validated in Ref.~\cite{Aguilar-Arevalo:2016ndq}.
We then introduced the simulated charge distribution on top of images in the background data sets, in order to obtain a realistic representation of the image noise.

\section{Image processing and event reconstruction}
\label{sec:event}

The image processing started with the determination of its pedestal value, corresponding to the dc offset introduced at the time of readout. 
The pedestal was estimated independently for each column of the image by a Gaussian fit to the distribution of pixel values in the column, and was then subtracted from every pixel.
After this first column-based equalization, the same procedure was applied to each row of the image, yielding a final image in which the distribution of the pixel values is centered at zero with standard deviation equal to the pixel noise ($\sigma_{\rm{pix}}$).
Hot pixels or defects were identified as recurrent patterns over images in the same data set and eliminated (``masked'') from the analysis (typically, $<$10\% of the pixels were removed by this procedure).

Ionization events were identified as clusters of contiguous pixels with values $>$4\,$\sigma_{\rm{pix}}$.
The energy of the event was estimated as the addition of the pixel values of the cluster.
For 1$\times$1 data, an additional cluster search was implemented for events with energies $<$10\,keV, for which the electron recoil track length is much smaller than the pixel size, and the distribution of charge on the pixel array is well described by the 2D Gaussian distribution arising from charge diffusion.
The algorithm is based on a moving window of 11$\times$11 pixels.
For a given window's position, the difference in log-likelihood ($\Delta LL$) between two hypotheses \textemdash\ the first of a 2D Gaussian distribution of charge on top of white noise, the second of only white noise \textemdash\ was calculated.
If the 2D Gaussian hypothesis was found more likely, the window was moved around to find the local $\Delta LL$ maximum, to properly center the event in the window.
Then a fit was performed from which the $x$-$y$ position, charge spread and energy ($E$) of the candidate ionization event were obtained as the best-fit values of the center ($\vec{\mu}$), standard deviation ($\sigma_{xy}$) and integral of the 2D Gaussian, respectively.
This clustering procedure has been previously validated in Ref.~\cite{Chavarria:2016xsi}, where it was shown that the requirement $\Delta LL$$<$$-22.5$ efficiently selects ionization events with a negligible contribution from readout noise.

\begin{figure}[t!]
	\centering
	\includegraphics[width=0.482\textwidth]{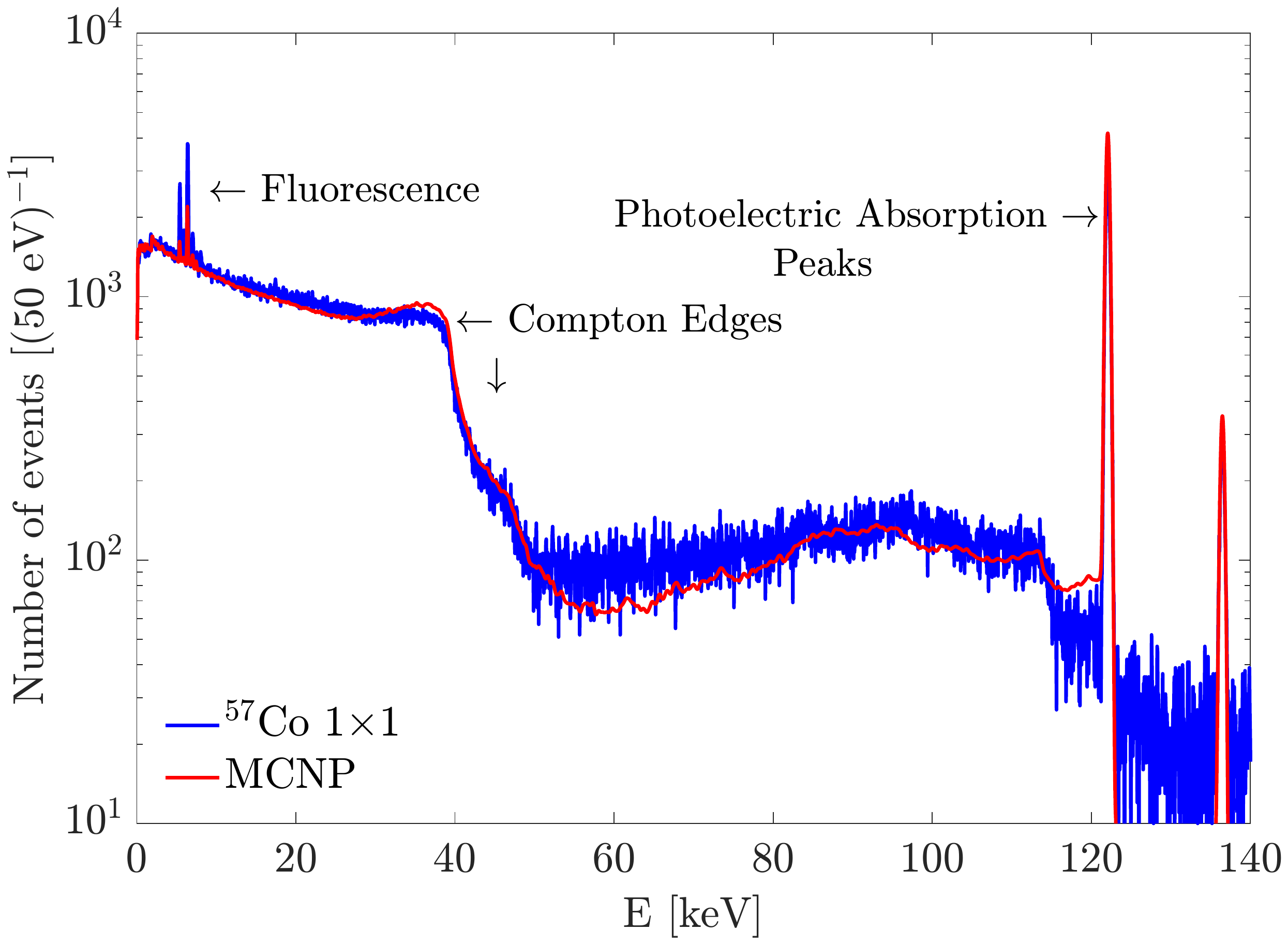}
	\caption{Spectrum observed in the 1$\times$1 data from the \cob\ source. The expectation from MCNP, with its amplitude normalized to the photoelectric absorption lines, is shown for reference. Characteristic spectral features are labeled.}
	\label{fig:mcnp_full}
\end{figure}

Figure~\ref{fig:mcnp_full} shows the observed electron-recoil spectrum in the 1$\times$1 data with the \cob\ source. Because of the high spatial resolution of the CCD, each event arises from a single $\gamma$-ray interaction. In addition to the spectral continuum from zero up to the Compton edges, other characteristic spectral features are evident, including the monoenergetic peaks from photoelectric absorption of the primary 122\,keV and 136\,keV $\gamma$ rays, and secondary x-ray fluorescence lines from the surrounding materials. The simulated spectrum with MCNP is shown for comparison, presenting generally a fair agreement to the data, with some underestimation of the fluorescence yields.

\begin{figure}[t!]
	\centering
	\includegraphics[width=0.482\textwidth]{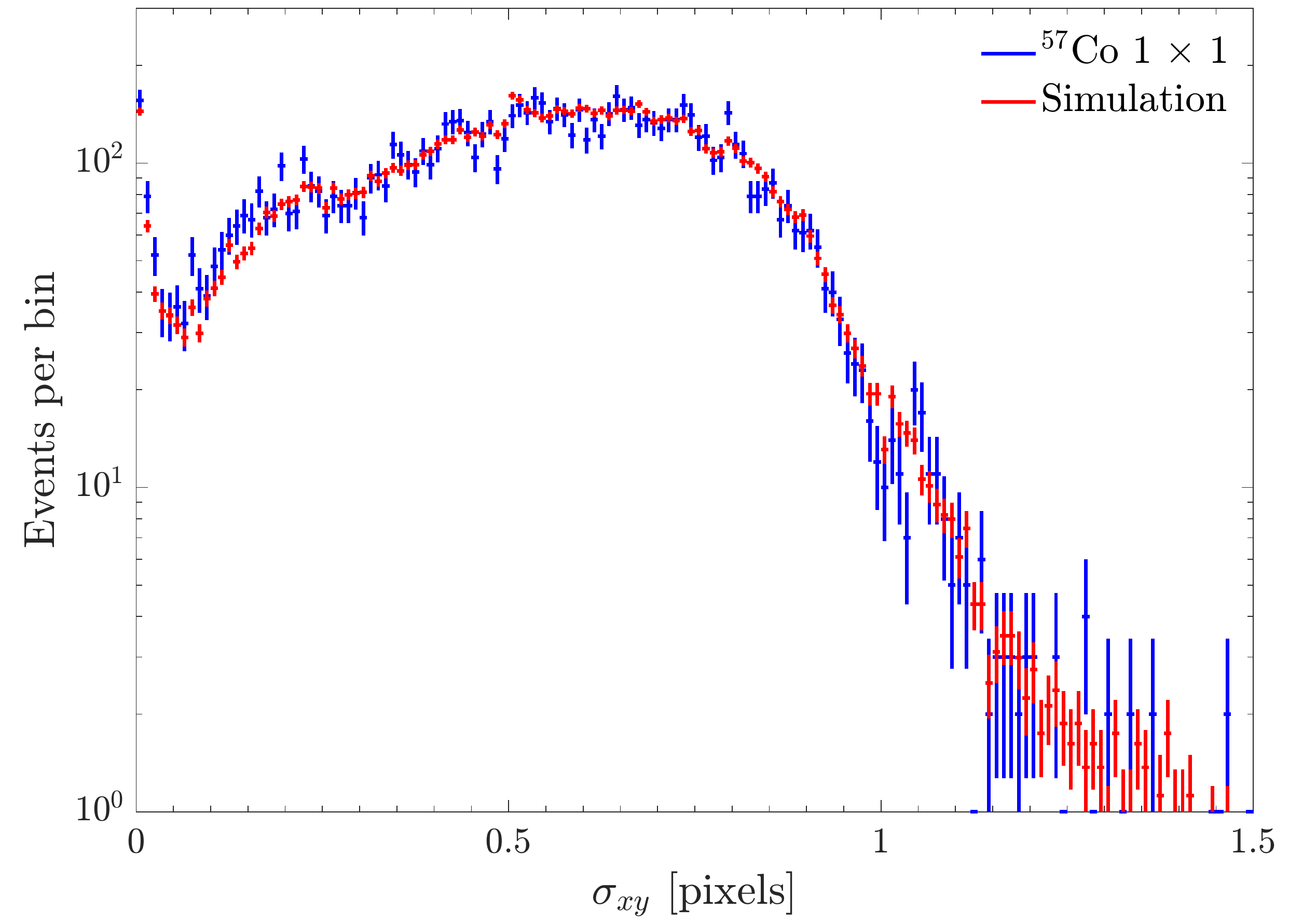}
	\caption{Lateral spread ($\sigma_{xy}$) of clusters in the \cob\ 1$\times$1 data with energies $<$1\,keV excluding readout noise, as obtained from the likelihood extraction described in Section~\ref{sec:event}. The result obtained by applying the same procedure to simulated events with a uniform distribution across the thickness of the device is shown for comparison.}
	\label{fig:diff}
\end{figure}

The scattering length in silicon of $\gamma$ rays with energies above 50\,keV is $>$1\,cm, much larger than the thickness of the CCD, leading to interactions that are distributed uniformly in depth.
Figure~\ref{fig:diff} presents the fitted $\sigma_{xy}$ of selected clusters with energies $<$1\,keV, including those with energies as small as 60\,eV.
The distribution is compared to the result obtained for simulated events with a uniform distribution in the 0--1\,keV energy range and a uniform spatial distribution across the thickness of the device.
The parameters of the diffusion model were tuned to events at higher energy, and are in good agreement with those inferred by scaling a previous calibration of 675\,$\mu$m-thick CCDs from SNOLAB~\cite{Aguilar-Arevalo:2016ndq}.
The close match between both distributions demonstrates that the recorded spatial distribution of low-energy clusters is consistent with the signal from Compton scattering, with a negligible contamination from surface events.

\section{Low-energy spectra}
\label{sec:spectra}

\begin{figure}[t!]
	\centering
	\includegraphics[width=0.482\textwidth]{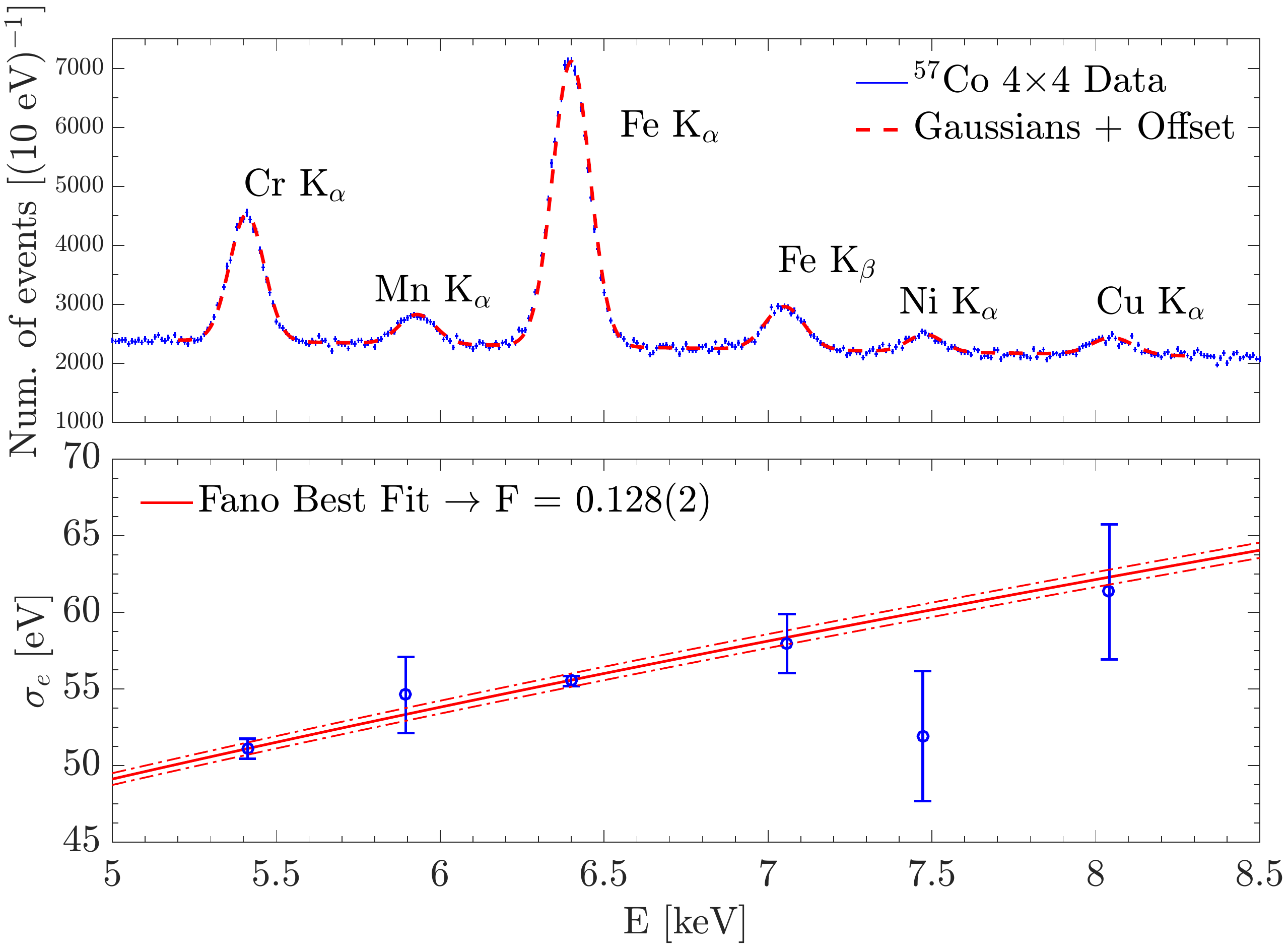}
	\caption{{\it Top:} Energy spectrum from the \cob\ source with Gaussian fits to the fluorescence lines used for calibration. {\it Bottom:} Energy dependence of the line width; the Fano factor was obtained from a fit with the energy resolution model described in Section~\ref{sec:spectra}.}
	\label{fig:fano}
\end{figure}

Low-energy spectra below 4\,keV were constructed from the 4$\times$4 data sets.
For each data set, the energy scale of the ionization signal was calibrated with in-run fluorescence x rays from the stainless steel chamber (Fig.~\ref{fig:fano}).
The linear response of the CCD has been demonstrated for signals as small as 10\,$e^-$~\cite{Aguilar-Arevalo:2016ndq}.

\begin{figure}[t!]
	\centering
	\includegraphics[width=0.482\textwidth]{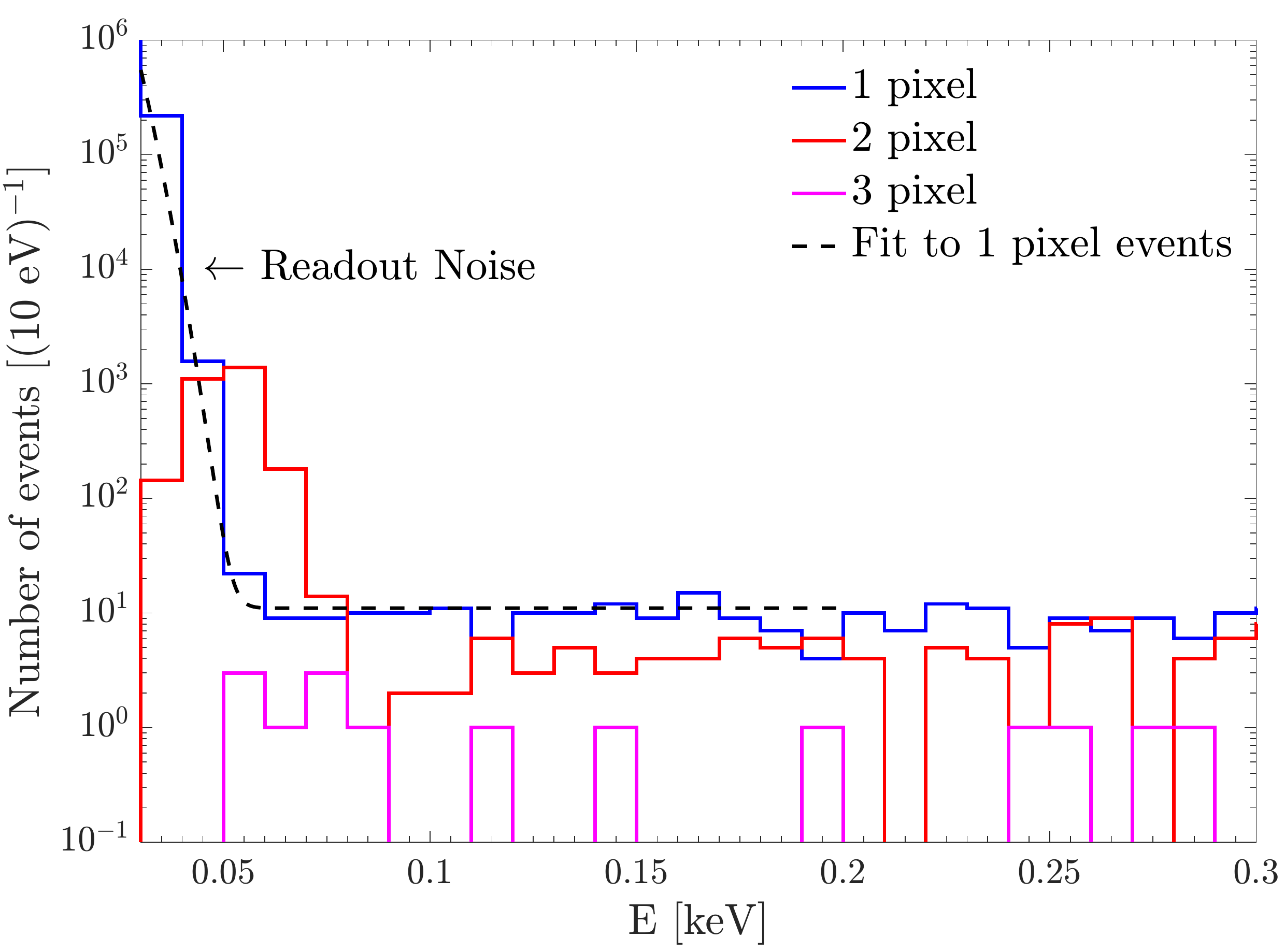}
	\caption{Energy spectra from the 4$\times$4 background data for clusters of different sizes. A Gaussian fit was performed to estimate the contribution of readout noise to the single-pixel spectrum.}
	\label{fig:bkgnd_pix}
\end{figure}

The energy threshold for this analysis was chosen to exclude readout noise. Figure~\ref{fig:bkgnd_pix} presents the spectrum of selected clusters with different number of pixels in all background data. The dashed line shows the result of a Gaussian fit to the single-pixel white noise, which demonstrates a negligible contribution of readout noise for single-pixel clusters above 60\,eV. Readout noise is still important for clusters with a larger number of pixels up to 80\,eV, becoming negligible beyond this point. Hence, to construct the final spectra, we consider only single-pixel events in the 60--80\,eV range and correct for the 95\%--90\% efficiency of this selection, as estimated from simulation. For energies $>$80\,eV, we consider all clusters without any correction, as the efficiency of clustering an event at these energies is already $>$99\,\% according to simulation.

\begin{figure*}[t!]
	\centering
	\includegraphics[width=0.75\textwidth]{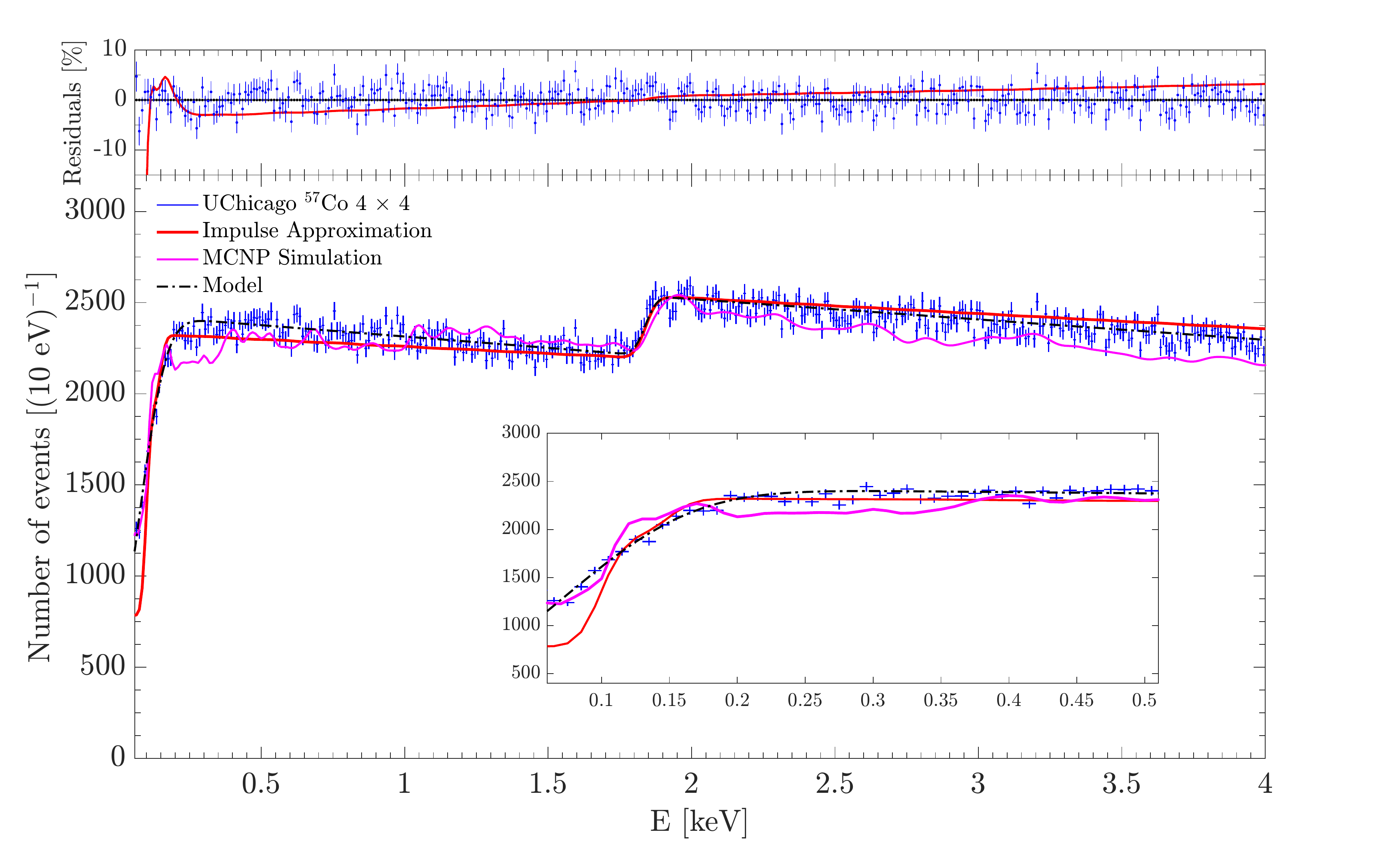}
	\caption{{\it Bottom}: Low-energy Compton spectrum from the \cob\ source. The predictions from the Impulse Approximation (IA) and MCNP are shown for comparison. The best fit to the model described in Section~\ref{sec:model} is presented by the dashed black line. {\it Inset:} Detail of the L step in the 60--500\,eV range. {\it Top:} Residuals after subtraction of the best-fit model from the data.}
	\label{fig:fit_co}
\end{figure*}

\begin{figure*}[t!]
	\centering
	\includegraphics[width=0.75\textwidth]{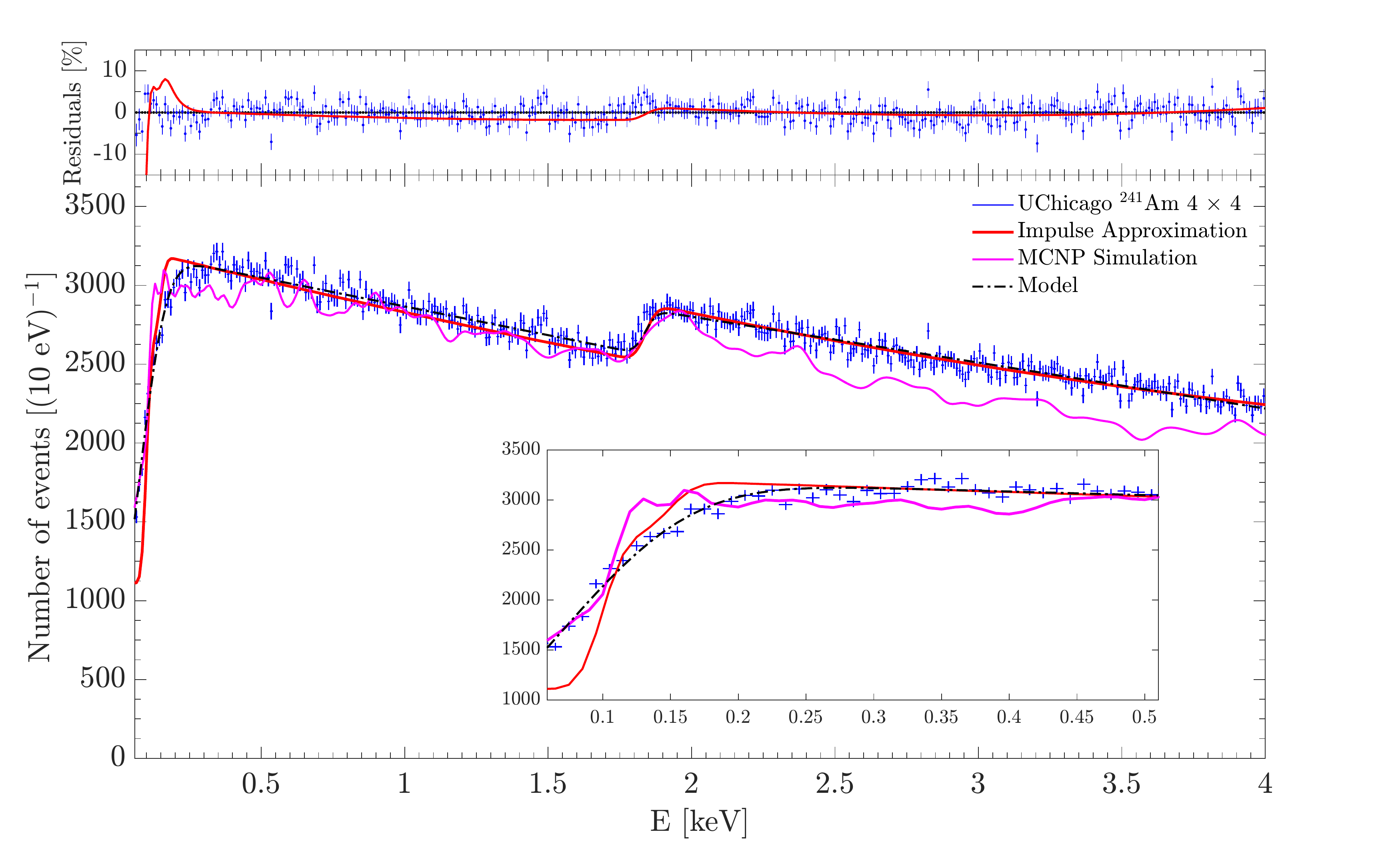}
	\caption{{\it Bottom}: Low-energy Compton spectrum from the \ame\ source. The predictions from the Impulse Approximation (IA) and MCNP are shown for comparison. The best fit to the model described in Section~\ref{sec:model} is presented by the dashed black line. {\it Inset:} Detail of the L step in the 60--500\,eV range. {\it Top:} Residuals after subtraction of the best-fit model from the data.}
	\label{fig:fit_am}
\end{figure*}

Spectra were measured for the \cob, \ame, and background data sets.
To remove the contribution of environmental backgrounds from the source spectra, the corresponding background spectra were scaled to the total exposure of the source spectra and then subtracted. Figures~\ref{fig:fit_co} and~\ref{fig:fit_am} show the final \cob\ and \ame\ Compton spectra in the 60\,eV--4\,keV range, respectively.
The data are compared to the predictions from IA and MCNP, scaled to match the right-hand side of the K step in the data.
The detector energy resolution was included in the predicted spectra, and was modeled as $\sigma_E^2$$=$$(12{\rm \,eV})^2$$+$$(3.8{\rm \,eV})FE$, where $F$$=$0.13, the Fano factor~\cite{PhysRev.72.26, *doi:10.1117/12.948704}, was directly estimated from the width of the observed fluorescence lines as shown in the bottom panel of Fig.~\ref{fig:fano}.
The constant term of $\sigma_E^2$ arises from the image pixel noise and was estimated from the analysis of monoenergetic low-energy events in simulated data sets.

Both IA and MCNP are able to correctly match the gross features of the Compton spectra at low energies, implying a satisfactory implementation of the underlying physics. Overall, IA provides a better match to the data than MCNP, but it fails to accurately reproduce the shape of the L-step feature (insets of Fig.~\ref{fig:fit_co} and~\ref{fig:fit_am}). This is unexpected, as the energy, amplitude and shape of the K step is consistent with the IA prediction. It is unlikely that the apparent decrease in resolution at the L step is due to the response of the detector. Our detailed calculations of the energy resolution, which consider the charge generated by all low-energy electrons~\cite{PhysRevB.22.5565} emitted in the Auger cascade~\cite{relax}, suggest that the Fano model should be valid at these energies. This is supported by the calibration of the detector with oxygen fluorescence x rays, which give a resolution of 21\,eV at $E_\gamma$$=$525\,eV. Thus, we interpret the apparent decreased resolution as originating from a softened L step in the electron recoil spectrum, which may occur if the theoretical assumption adopted by IA that each atomic shell may be treated independently does not strictly hold beyond the K shell.

\section{Model of Compton spectra at low energies}
\label{sec:model}

As is evident from Fig.~\ref{fig:fit_co} and~\ref{fig:fit_am}, and expected from IA (Fig.~\ref{fig:Biggs}), the Compton spectra at low energies is rather generic, with the position of the steps determined by the atomic shells of the target and the slope of the spectrum between the steps being approximately constant. We have compared the prediction from IA for energies $<$4\,keV to a piecewise function constructed from first-order polynomials bounded by the atomic binding energies. With the appropriate choice of parameters, the function agrees to better than 0.5\% with IA for a wide range of $\gamma$-ray energies for which the Compton scattering cross section is significant.

Motivated by this result, we propose to parametrize Compton spectra in the energy range 60\,eV--4\,keV with an expression of the form

\begin{equation}
f(E) = A \times
\begin{cases} 
	a_1(E-E_K) + 1 & E \ge E_K \equiv E_{10}  \\
	a_2(E-E_K) + b_2 & E_{L} \le E < E_K \\
	b_3 &  E < E_{L}, \\
\end{cases}
\label{eq:piecewise}
\end{equation}
with an additional Gaussian resolution term $\sigma_L$ that applies only for $E$$<$0.5\,keV to smooth the L-step feature. Although the IA predicts two distinct $L_1$ and $L_{2,3}$ steps, we include a single step at an effective energy $E_L$. With the appropriate choice of $\sigma_L$ and $E_L$, this definition significantly improves the description of the data by the model over the result from IA presented in Section~\ref{sec:spectra}.

To simplify the model further, we introduce the following approximation:

\begin{equation*}
b_3 =  \frac{Z-10}{Z-2}[b_2+a_2(E_L-E_K)],
\end{equation*}
which defines the relative amplitude of the spectrum before and after the $L$ step as the ratio of the number of target electrons that contribute to the signal below and above $E_L$. After imposing this constraint, the number of free parameters of the proposed parametrization decreases to six: the amplitude of the K step ($h=1-b_2$), the slopes to the right and left of the K step ($a_1$ and $a_2$), the parameters defining the shape of the L step ($\sigma_L$ and $E_L$), and the overall normalization of the spectrum ($A$).

\begin{figure}[t!]
	\centering
	\includegraphics[width=0.482\textwidth]{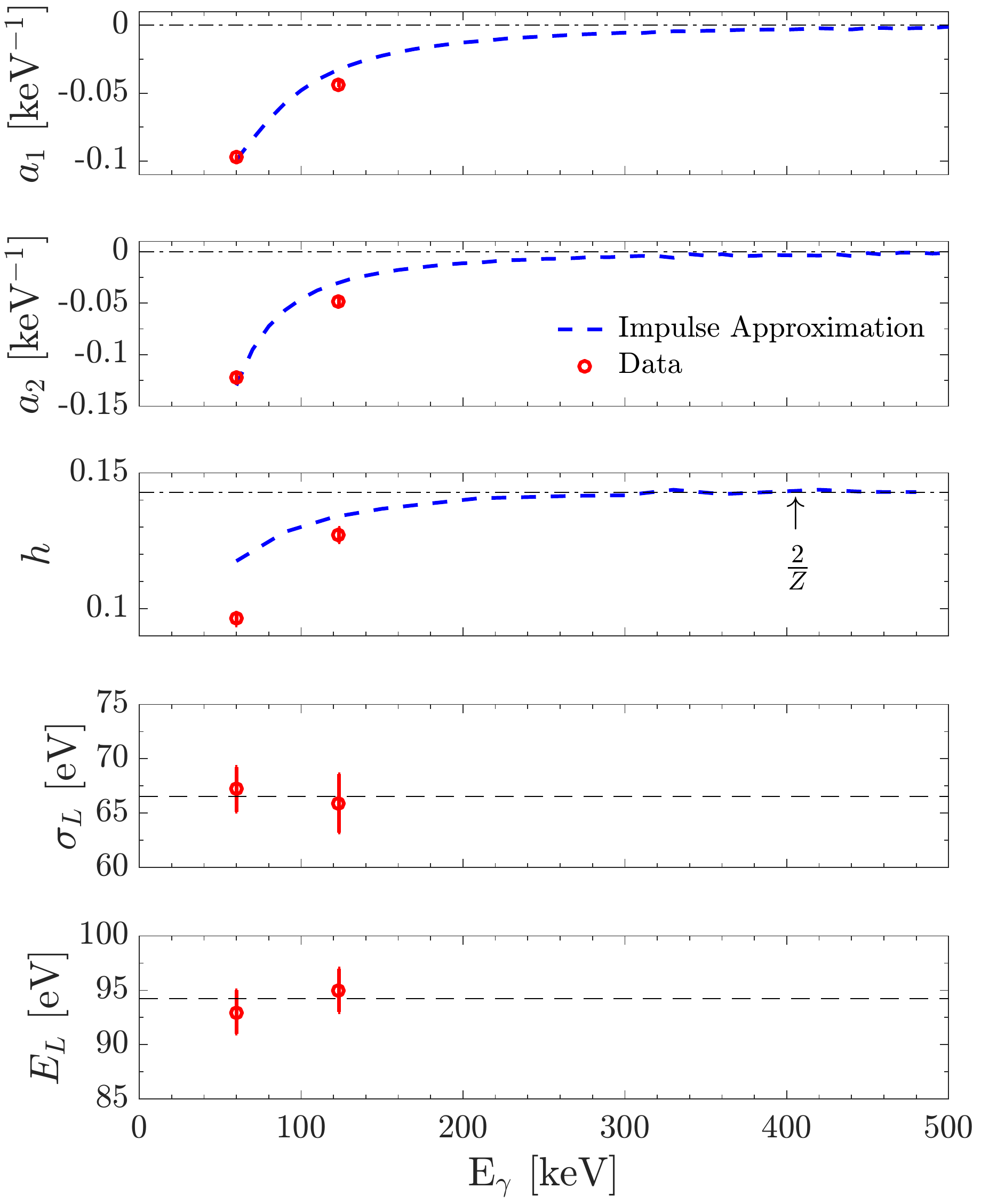}
	\caption{Dependence of the five model parameters that determine the Compton spectrum at low energies, i.e, $a_1$, $a_2$, $h$, $\sigma_L$ and $E_L$, on the incident $\gamma$-ray energy. The dashed blue line in the top three panels shows the prediction from IA, while the dashed black line of the bottom two panels is the mean value of both measurements.}
	\label{fig:param}
\end{figure}

The best fit to the data with our model (including the detector response presented in Section~\ref{sec:spectra}) is shown by the dashed black line in Fig.~\ref{fig:fit_co} and~\ref{fig:fit_am} with the residuals in the top panel, which demonstrate an agreement to better than 5\% throughout the full energy range. Figure~\ref{fig:param} shows the best-fit values as a function of $\gamma$-ray energy for the five parameters that determine the shape of the spectrum. For $a_1$, $a_2$ and $h$, we included the prediction by IA. Our observations are consistent with the expectations from IA: i) the magnitude of the slope of the spectrum is inversely proportional to $E_\gamma$, and ii) the value of $h$ asymptotically approaches the fraction of electrons in the K shell, i.e., $2/Z$. The best-fit values of $\sigma_L$$\sim$65\,eV and $E_L$$\sim$95\,eV are consistent between the \cob\ and \ame\ data, with no dependence on $E_\gamma$.

Because of the linear nature of Eq.~\ref{eq:piecewise} and the constant values of $\sigma_L$ and $E_L$, the addition of multiple functions corresponding to incident $\gamma$ rays of different energies would also be accurately described by the same function with the appropriate choice of average values for the parameters. Therefore, the proposed parametrization should be a good description for the Compton background at low energies for any energy distribution of the incident $\gamma$-ray flux.

\section{Conclusions}
\label{sec:conclusion}

A dominant source of environmental background in direct searches for low-mass dark matter particles are the energy deposits by small-angle Compton scattering of $\gamma$ rays.
We performed detailed measurements of Compton spectra between 60\,eV and 4\,keV in silicon, which demonstrate the capability of the CCDs employed in DAMIC to reliably resolve spectral features down to the experiment's current threshold.
We report, for the first time, spectral features associated with the atomic structure of the target, and present a general parametrization to describe the Compton spectrum at low energies for any energy distribution of the incident $\gamma$-ray flux.
The model is based on the theoretical prediction of the Impulse Approximation modified for energies $<$0.5\,keV, where the theory fails to describe the data.
The inadequacy of the theoretical prediction at the lowest energies stresses the importance of precise experimental studies to characterize the backgrounds for low-mass dark matter searches.
Our results are directly applicable to background estimates for the DAMIC experiment, as well as for other direct searches for dark matter that employ silicon detectors~\cite{Agnese:2016cpb}.

\begin{acknowledgments}
This work has been supported by the Kavli Institute for Cosmological Physics at the University of Chicago through Grant No. NSF PHY-1125897 and an endowment from the Kavli Foundation, and by Grant No. NSF PHY-1506208.
\end{acknowledgments}

\bibliography{Gamma.bib}

\end{document}